\documentclass[preprint,12pt,authoryear]{elsarticle}

\usepackage{amssymb}
\usepackage{amsmath}

\usepackage{epsfig}
\usepackage{geometry}
\usepackage{caption,keyval}
\usepackage{gensymb}
\usepackage{multirow}
\usepackage{latexsym}
\usepackage{amsfonts}
\usepackage{booktabs}
\usepackage{subcaption}
\usepackage{tabularx}
\usepackage{graphicx}
\usepackage[colorlinks=true, pdfborder={0 0 0}, linkcolor=blue]{hyperref}
\usepackage[detect-all]{siunitx}
\usepackage{xcolor}

\journal{Chemical Engineering Science}

\begin{document}

\begin{frontmatter}



\title{Predicting the distribution of
yield-stress fluids in branched pipe manifolds} 


\author[inst1]{Elliott Sutton}
\author[inst2]{Waldo Rosales Trujillo}
\author[inst2]{Adam Kowalski}
\author[inst1]{Cláudio P. Fonte\corref{cor1}}
\author[inst3,inst4]{Anne Juel\corref{cor2}}

\affiliation[inst1]{organization={Department of Chemical Engineering, The University of Manchester},
            addressline={Oxford Road}, 
            city={Manchester},
            postcode={M13 9PL},             
            country={UK}}

\affiliation[inst2]{organization={Unilever R\&D, Port Sunlight Laboratory},
            addressline={Wood Street},
            city={Wirral},
            postcode={CH62 4UY},
            country={UK}}
            
\affiliation[inst3]{organization={Department of Physics \& Astronomy, The University of Manchester},
            addressline={Oxford Road}, 
            city={Manchester},
            postcode={M13 9PL}, 
            country={UK}}
            
\affiliation[inst4]{organization={Manchester Centre for Nonlinear Dynamics, The University of Manchester},
            addressline={Oxford Road}, 
            city={Manchester},
            postcode={M13 9PL}, 
            country={UK}}
            
\cortext[cor1]{E-mail: claudio.fonte@manchester.ac.uk}
\cortext[cor2]{E-mail: anne.juel@manchester.ac.uk}

\begin{abstract}
We develop a one-dimensional network model to predict the steady-state distribution of yield-stress fluids in branched pipe manifolds under wall-slip conditions. The model accounts for major friction losses between junctions and incorporates wall slip through a power-law relation calibrated independently via capillary rheometry. Predictions from the model are validated against both bench-scale experiments and fully resolved computational fluid dynamics simulations, showing excellent agreement across a range of flow conditions. Our results demonstrate that wall slip strongly influences the uniformity of fluid distribution by modifying the relative resistance between outlet branches. Furthermore, we show that the problem can be inverted: measured distribution profiles can be used to estimate slip parameters, offering a practical method for slip characterisation without pressure measurement. This modelling framework is computationally inexpensive, robust, and adaptable to various network configurations, making it a valuable tool for the design and analysis of industrial manifold systems involving viscoplastic fluids.

\end{abstract}


\begin{highlights}
    \item A reduced-order network model accurately predicts yield-stress fluid distribution in pipe manifolds.
    \item Wall slip is incorporated via a power-law model calibrated using capillary rheometry.
    \item Model predictions show excellent agreement with both experiments and 3D flow simulations.
    \item Wall slip enhances distribution uniformity by reducing resistance disparities between branches.
    \item The manifold system can be used to extract slip parameters without direct pressure measurements.
\end{highlights}

\begin{keyword}


Yield-stress fluids \sep
Wall slip \sep
Network modelling \sep
Pipe manifolds \sep
Viscoplastic flow \sep
Flow distribution \sep
Slip characterisation
\end{keyword}

\end{frontmatter}



\section{Introduction}
\label{section:introduction}
In many industrial filling operations, fluids are delivered from the process to a series of filling nozzles via a pipe manifold located at the end of the manufacturing line. The challenge for designers of such processes is to devise a manifold configuration that can achieve the desired distribution (typically uniform) amongst many outlets. Maldistribution arises from dissimilar resistances to flow in the outlet branches \citep{Miller:1978, Yazici-et-al:2024}, as well as dynamical effects that arise due to inertial \citep{Chen-Sparrow:2009}, thermal \citep{Baikin-Taitel-Barnea:2011,Aka-Narayan:2022}, or elastic instabilities \citep{Varshney-et-al:2016} in different sections of the pipeline such as bends and bifurcations.

Previous studies have focused on Newtonian fluid flow through manifolds and have successfully applied network model methodologies to predict fluid distribution \citep{Miller:1978, Majumdar:1980,Kee-et-al:2002}. The distribution profile in the manifold outlets depends on the resistance to flow throughout the different paths of the network, determined by the total pressure drop across each branch. The Hagen-Poiseuille equation provides an analytical description of major friction losses in straight pipe segments of a manifold, and loss factors describe minor friction losses in junctions and bends; minor losses are empirically derived factors specific to particular flow configurations. Additional loss mechanisms cause minor losses in Newtonian laminar flows and arise due to inertia, which forms secondary flow structures in bends \citep{Dean:1928} and bifurcations \citep{Chen-Rowley-Stone:2017}.

By contrast with the vast literature on Newtonian fluids, the flow of non-Newtonian fluids through pipe manifolds is sparse. There have been some attempts to use network models to describe distribution systems, such as the work by \citet{Brod:2003} for polymer melts. They aimed to minimise residence times by selecting optimal tube diameters for a given system design and imposed pressure gradient. They report that ideal designs maintain a uniform apparent wall shear rate throughout the network. Network models can also describe the flow through porous media, which distribute fluid through a network of pores. \citet{Fraggedakis-Chaparian-Tammisola:2021} could accurately predict the first fluidised path of viscoplastic fluid through a porous medium utilising a Dijkstra algorithm, avoiding computationally expensive flow simulations. Similarly, shear-thinning \citep{Rodriguez-Goyeau:2021} and viscoplastic \citep{Balhoff-Thompson:2004, Balhoff-et-al:2012, Liu-et-al:2019} fluid flows through porous media were described by network models, reporting modifications to Darcy's law. These models are akin to the network model of \citet{Miller:1978}, calculating flow resistance to predict preferential routes and flux through a medium. To the best of our knowledge, no studies have directly investigated yield-stress fluid flow through pipe manifolds.

Yield-stress fluids are prone to slipping on solid surfaces. Wall slip, when present, drastically changes the behaviour of yield-stress fluid flow by reducing friction between the fluid and a solid boundary and permits displacement from imposed stresses insufficient to induce yielding \citep{Kamdi-Orpe-Kumaraswamy:2021}. As such, a network model based on frictional resistance should account for this phenomenon, though previous models have typically neglected it. Many yield-stress fluids exhibit apparent wall slip due to the formation of a thin lubrication layer near the boundary, typically with a thickness on the order of 10~nm \citep{Seth-Cloitre-Bonnecaze:2008, Zhang-et-al:2017}. This layer, composed of the continuous phase or solvent, has a significantly lower viscosity than the bulk material. Consequently, the bulk material appears to slip at the wall, although this effect arises from flow within the low-viscosity lubrication layer rather than actual motion at the solid boundary. In soft jammed suspensions, \citet{Pemeja-et-al:2019} identified two distinct slip regimes, depending on the magnitude of the wall shear stress: a \emph{linear regime}, in which the slip velocity is directly proportional to the applied shear stress, and a \emph{nonlinear regime}, in which the slip velocity scales quadratically with stress. In the linear regime, dissipation is described by a Couette flow within the continuous-phase lubrication layer, while the particulate bulk behaves as a rigid solid sliding over it. The formation of this lubrication layer can be attributed to repulsive/attractive forces between the occlusions of the material and solid walls, and to the inability of the material to retain the bulk jammed structure in the immediate vicinity of the wall due to geometric confinement \citep{Zhang-et-al:2017}. The elastohydrodynamic lubrication model proposed by \citet{Meeker-Bonnecaze-Cloitre:2004a,Meeker-Bonnecaze-Cloitre:2004b} accounts for the nonlinear regime, wherein hydrodynamic lift depletes further the near-wall region of suspended particles, setting the effective thickness of the lubrication layer. This thickness is governed by a balance between the hydrodynamic lift force, the osmotic pressure of the bulk suspension, and the elastic deformation of the particles. Both slip mechanisms become more prominent at low stresses, below the yield stress. Above the yield stress, apparent slip persists but gradually becomes negligible as the material's microstructure is disrupted under increasing deformation. This transition has been observed in both steady-shear \citep{Buscall-et-al:1993} and oscillatory-shear \citep{Walls-et-al:2003} measurements.

    Modelling the effects of wall slip on flow has been an area of interest in the literature, with particular attention paid to predicting velocity profiles in complex geometries. Analytical expressions predicting flow in simple geometries are straightforward to obtain, and some exist for non-trivial geometries such as curved channels \citep{Cox-Taghavi:2025}. For more complex systems, numerical analysis is required. Numerous studies have implemented wall slip into augmented Lagrangian and finite element methods \citep{Roquet-Saramito:2008,Chaparian-Tammisola:2021,Muravleva:2021}, enabling numerical prediction of fluid velocities and pressure drops in non-trivial geometries, such as non-circular ducts and porous media. These studies have been very successful in predicting fluid behaviour at solid boundaries; for instance, \citet{Chaparian-Tammisola:2021} calculated the resistance to flow in two-dimensional porous-media geometries by resolving the entire flow field numerically with their augmented Lagrangian method, thereby probing the effect of wall slip on pressure drop. Such methods can, however, be highly computationally expensive. To mitigate this, \citet{Roquet-Saramito:2008} implemented adaptive mesh refinement to improve prediction of the stick–slip transition while reducing computational demands. Accurate, lower-fidelity methods, such as network models, can offer further improvements in computational efficiency when the entire flow field need not be resolved and are, therefore, an accessible alternative to full computational fluid dynamics simulations. Credible numerical modelling of real-world systems nevertheless requires characterisation of a fluid’s slip properties. Capillary and rotational rheometry \citep{Meeker-Bonnecaze-Cloitre:2004a,wilms-et-al:2021} and flow visualisation \citep{Perez-Gonzalez-et-al:2012,Daneshi-et-al:2019} have been used successfully to measure slip, but the subject remains under active investigation, and established protocols depend highly on the nature of the yield-stress fluid.

The lack of literature characterising the flow of yield-stress materials through pipe manifolds and the inability of process engineers to accurately predict the fluid distribution from a manifold design is a significant hindrance to industry. Furthermore, computational fluid dynamics simulations (CFD) of yield-stress fluid flow in such large geometries converge notoriously slowly \citep{Saramito-Wachs:2017} and are prohibitively expensive. Thus, in this study, we develop a reduced-order network model capable of modelling the flow with a fraction of the computational cost and convergence time. We test its capabilities against experimental data and fully resolved CFD simulations. The remaining manuscript is organised as follows. In Section~\ref{section:Methodology}, we describe the numerical and experimental approaches used in this study. Section~\ref{section:results} presents the results, comparing our model's predictions with both experimental and computational data. We also examine how slip affects the distribution produced by a manifold. Finally, our conclusions are summarised in Section~\ref{section:conclusions}.

\section{Methodology}
\label{section:Methodology}

This section presents the formulation of the network model, followed by the experimental and computational methodologies used for its validation.

\subsection{Network Model Formulation}
\label{section:network model}

In the proposed model, manifold geometries comprising one inlet and an arbitrary number of outlet branches are represented as a one-dimensional flow network. Each junction, including the inlet and outlet points and the bifurcation points between channels, is treated as a node in the network. These nodes are connected by straight, circular channels corresponding to the physical segments of the manifold. The flow is assumed to be steady and fully developed in each segment, and mass conservation is enforced at each node $i$ according to
\begin{equation}
    \sum_{j} \dot{m}_{i,j} = 0 \;\;,
    \label{eqn: mass conserv}
\end{equation}
where $\dot{m}_{i,j}$ is the mass flow rate from node $i$ to all its neighbouring nodes $j$.

We describe the rheology of the liquids of interest for this work using the Herschel–Bulkley model \citep{Herschel-Bulkley:1926},
\begin{equation}
    \label{eqn:HB}
    \hat{\tau} = \tau_0 + K\hat{\dot{\gamma}}^{n} \;\;,
\end{equation}
where $\hat{\tau}$ is the shear stress, $\tau_0$ the yield stress, $K$ the consistency index, $n$ the flow index, and $\hat{\dot{\gamma}}$ the shear rate under steady, unidirectional shear. A circumflex is used to denote dimensional variables where a corresponding non-dimensional form is also introduced additionally in the manuscript.

Two types of pressure losses may arise in the network: (i) major losses due to friction between the fluid and the channel walls, and (ii) minor losses, originating from junctions and bends. Following the results of \citet{Sutton-et-al:2022} on the flow of viscoplastic fluids in pipe bends, minor losses can be considered negligible when the Reynolds number is $Re \lesssim 100 $. For a Herschel–Bulkley fluid flowing through a circular channel, the Reynolds number from nondimensionalisation of Cauchy's momentum equation (see \ref{appendix:model nondimensionalisation}) is
\begin{equation}
    \label{eqn:HB Re}
    Re = \frac{\rho \hat{\bar{u}}^{2-n} D^n}{K} \;\;,
\end{equation}
where $\rho$ is the density of the fluid, $\hat{\bar{u}}$ the average cross-sectional velocity, and $D$ the diameter of the channel. Across all tested conditions, the branch Reynolds numbers computed from Eq.~\eqref{eqn:HB Re} remained below 100, so bend/junction minor losses are negligible in this laminar viscoplastic regime and are omitted. These conditions are representative of typical operating conditions of practical interest.

An analytical momentum conservation equation is derived for the fully developed laminar flow of a Herschel--Bulkley fluid with wall slip in a straight circular channel. The detailed derivation and nondimensionalisation are presented in \ref{appendix:model nondimensionalisation}. The resulting nondimensional momentum equation is
\begin{equation}
    \label{eqn:HB SP Slip}
    S \tau_\mathrm{w}^{\beta} + \frac{n\left(\tau_\mathrm{w} - B \right)^{\frac{n+1}{n}}}{2\tau_\mathrm{w}^3}\frac{ \left(n+1\right)\left(2n+1\right)\tau_\mathrm{w}^2 + 2n\left(n+1\right)\tau_\mathrm{w} B + 2n^2 B^2}{\left(n+1\right)\left(2n+1\right)\left(3n+1\right)} = 1
\end{equation}
for $\tau_\mathrm{w} > B$, and reduces to
\begin{equation}
    \label{eqn:HB SP Only Slip}
    S \tau_\mathrm{w}^\beta = 1
\end{equation}
for $\tau_\mathrm{w} \leq B$, where $\tau_\mathrm{w}$ is the wall shear stress. The Bingham number, $B$, is defined as
\begin{equation}
    \label{eqn:Bingham number}
    B = \frac{\tau_0}{K\left(\hat{\bar{u}}/D\right)^n} \;\;,
\end{equation}
and is the ratio of the yield stress to viscous stresses acting on the fluid. The nondimensional slip number, $S$, is given by
\begin{equation}
    \label{eqn:slip number}
    S = \frac{\alpha K^\beta \hat{\bar{u}}^{\beta n-1}}{D^{\beta n}} \;\;,
\end{equation}
where $\alpha$ and $\beta$ are empirical parameters from the wall slip, $\hat{u}_\mathrm{s}$, constitutive law \citep{Zhang-et-al:2017}
\begin{equation}
    \label{eqn:slip law}
    \hat{u}_\mathrm{s} = \alpha \hat{\tau}_\mathrm{w}^\beta \;\;,
\end{equation}
and quantifies the contribution of wall slip to the dynamics of the flow, i.e. wall slip effects increase with $S$, and no-slip conditions occur with $S=0$. Finally, the pressure drop, $\hat{p}_i-\hat{p}_j$, in the channel connecting nodes $i$ and $j$ is calculated from the wall shear stress obtained from the solution of Eqs.~\ref{eqn:HB SP Slip} and \ref{eqn:HB SP Only Slip} as
\begin{equation}
    \label{eqn:pressure drop}
    \hat{p}_i-\hat{p}_j = 4 \hat{\tau}_\mathrm{w} \frac{L_{i,j}}{D_{i,j}},
\end{equation}
where $L_{i,j}$ and $D_{i,j}$ are the length and local diameter of the channel.

The mass conservation and pressure continuity equations [Eqs.~\eqref{eqn: mass conserv} and \eqref{eqn:pressure drop}] enforced at each node of the network define a system of coupled nonlinear equations, which we solve numerically. For Newtonian fluids, the problem is linear and amenable to matrix inversion. However, the Herschel–Bulkley model introduces nonlinearity, which we resolve using the \texttt{fsolve} function in \textsc{Matlab}. Converged solutions required initial guesses sufficiently close to the solution, for which we found the Newtonian solution with constant viscosity $\mu = K$ and $n=1$ to be effective. Boundary conditions include a prescribed inlet flow rate and atmospheric pressure at the outlets ($p_{\mathrm{out}} = 0$).

\subsection{Experimental Methodology}
\label{section:experimental methodology}
\subsubsection{Materials and Rheometry}
\begin{figure}[t]
\centering
\includegraphics[scale=1]{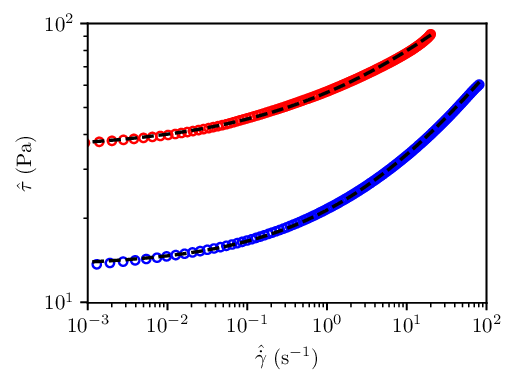}
\caption{Ramp-down steady-shear flow curves of Carbopol (blue circles) and the emulsion (red circles) using profiled parallel plates with a gap size of 1.55~mm. The black dashed curves represent the best fitting to the Herschel-Bulkley constitutive model [cf. Eq.~\eqref{eqn:HB} for the expression and Table~\ref{tab:material rheology} for the model parameters].}
\label{fig:rheological slip}
\end{figure}

We used a 1~g~L$^{-1}$ Carbopol Ultrez-21 microgel (Lubrizol) and a commercially available food emulsion (mayonnaise) as our test fluids. We prepared the Carbopol solution following the procedure outlined by \citet{Garg-et-al:2021}. No additional preparation was performed on the emulsion prior to use, other than allowing it to equilibrate to the laboratory ambient temperature. We measured the density of each material using a pipette to measure 10~mL of the fluid, which we weighed with a Mettler Toledo XSR225 balance. Stress‑controlled steady‑shear measurements were conducted using an Anton Paar MCR‑302 rheometer equipped with a parallel‑plate geometry (diameter \(50~\mathrm{mm}\)) at a controlled laboratory temperature of \(20~^\circ\mathrm{C}\). Profiled plates were used to minimise wall slip. The resulting flow curves for each material (Fig.~\ref{fig:rheological slip}) were fitted to the Herschel--Bulkley model [cf. Eq.~\eqref{eqn:HB}] to determine the model parameters. 

\begin{figure}[t!]
    \centering
    \includegraphics[scale=1]{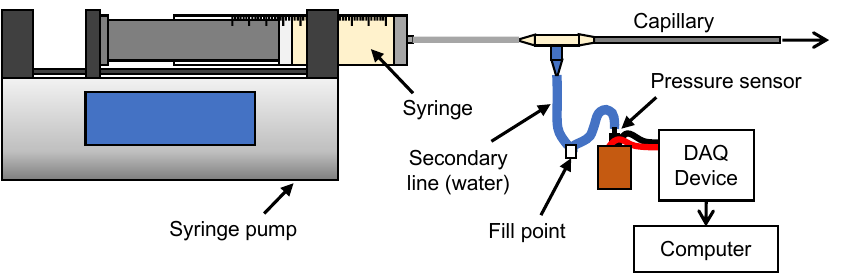}
    \caption{A schematic diagram of the capillary rheometer used to measure the slip properties of the yield-stress materials and calibrate the pressure measurement system. The setup features a syringe pump which controls the flux of material flowing through a capillary and a pressure sensor mounted upstream.}
    \label{fig:capillary rheometer schematic}
\end{figure}

The parameters $\alpha$ and $\beta$ of the slip model [Eq.~\eqref{eqn:slip law}] are acquired independently of our main experimental rig using capillary rheometry. A schematic representation of the capillary rheometer setup is shown in Fig.~\ref{fig:capillary rheometer schematic}. In this setup, fluid flow is induced through a capillary tube (length $L = 130.7$~mm and diameter $D=1.55$~mm) using a KD-Scientific Legato 200 syringe pump. We recorded the pressure drop across this capillary tube for a range of flowrates. A 5~psi Honeywell piezoresistive pressure sensor is mounted upstream of the capillary via a fluid-filled secondary line, using a T-junction to achieve a flush connection to the mainline without perturbing the flow. The sensor provides an analogue voltage signal that is amplified and converted to a digital signal by a National Instruments USB-6251 multi-functional data acquisition (DAQ) device and recorded by a computer. The recorded voltage signal is then converted to a pressure reading using a calibration curve acquired by attaching the sensor to an Elveflow OB1-4 pressure controller. We verified the accuracy of the pressure measurement in the capillary rheometer with a Newtonian fluid and present these results in \ref{appendix:pressure measurement system}. Once we confirmed the accuracy of this setup, we performed the measurements with our test fluids, fitting the slip parameters by matching the pressure drop predicted by Eq.~\eqref{eqn:HB SP Slip} and Eq.~\eqref{eqn:HB SP Only Slip} to the data. We provide additional discussion on this fit in \ref{section:wall slip}. The density, rheological, and slip parameters of the test fluids are summarised in Table~\ref{tab:material rheology}.

\begin{table}[ht]
    \centering
    \caption{Density, and rheological and slip parameters for the test materials at 20~$^\circ$C.}
    \begin{tabular}{l c c c c c c c}
        \toprule
        Materials & $\rho$ (kg m$^{-3}$) & $\tau_0$ (Pa) & $K$ (Pa s$^{n}$) & $n$ & $\alpha$ ($\mathrm{Pa}^{-\beta}\, \mathrm{m}\, \mathrm{s}^{-1}$) & $\beta$ (--) \\
        \midrule
        Carbopol & 1010 & 13.5 & 7.94 & 0.41 & $1.34\times10^{-5}$ & 1 \\
        Emulsion & 938 & 35.2 & 21.4 & 0.32 & $1.09\times10^{-6}$ & 2 \\
        \bottomrule
    \end{tabular}
    \label{tab:material rheology}
\end{table}

\subsubsection{Experimental Rig}
\begin{figure}[t!]
    \centering
    \includegraphics[scale=1]{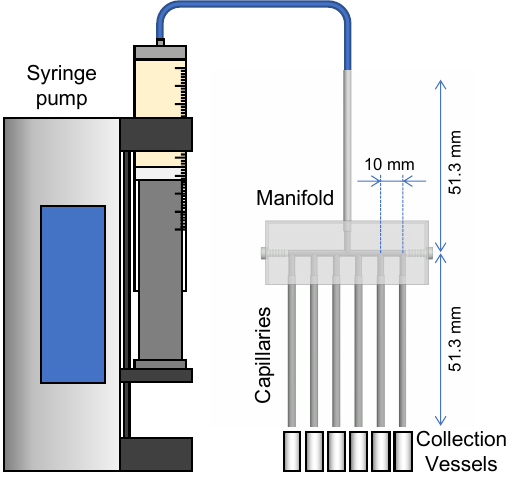}
    \caption{A schematic diagram of the experimental rig. A syringe pump delivers a constant flux of material to the manifold which divides the flow from one inlet to six outlets. Screws seal off the ends of the main manifold pipe.}
    \label{fig:experimental rig schematic}
\end{figure}

Fig.~\ref{fig:experimental rig schematic} shows a schematic diagram of the experimental rig used in this work. It consists of a manifold with six outlets ($N=6$) and an inlet positioned between the third and fourth outlets; Fig.~\ref{fig:experimental rig schematic} also includes a schematic of the manifold, indicating its dimensions. The manifold geometry is symmetric about the inlet, with three outlets on each side. This symmetry serves two purposes: in the experiments, deviations from a symmetric flow rate distribution can reveal manufacturing imperfections or the onset of symmetry-breaking flow instabilities; in the model, asymmetries in the predicted profile may indicate shortcomings in the modelling assumptions or numerical implementation. We fabricated the manifold by machining semi-circular channels ($D=1.55$~mm) into two aluminium blocks using a micro-miller. The inlet and outlets of the manifold are connected to stainless steel capillaries with an inner diameter of 1.55~mm and an outer diameter of 1.80~mm. We re-drilled the connection points on the manifold to match the outer diameter of the capillaries, permitting a consistent inner diameter between the manifold and capillaries without expansions or contractions. The manifold is assembled by clamping the two halves together with the capillaries in place under significant compression to prevent leaks. A syringe pump provided a constant flow rate of material into the pipe network. We determined the distribution profile from the manifold by capturing the output material from each outlet branch in collection vessels and recording the change of mass of the vessels after a fixed test duration, $\dot{m}_{\mathrm{out,\,}i}$, with an average of four trials taken. We present this mass in Section~\ref{section:results} as a fraction of the total input to the system, 
\begin{equation}
    \phi_i = \dot{m}_{\mathrm{out,\,}i} / \sum_{i=1}^{N} \dot{m}_{\mathrm{out,\,}i} \;\;.
\end{equation}
A minimum of 20~mL of material was passed through the manifold in each run to sufficiently reduce measurement errors in the outlet mass. Each trial lasted at least seven minutes to minimise the influence of flowrate ramp-up and ramp-down due to the syringe pump at the start and end of each run, which are on the order of seconds.

\subsection{Fully Resolved Flow Simulations}
\label{section:numerical methodology}
CFD simulations were used to support the experimental investigation. Three-dimensional simulations were performed using the open source finite-volume solver \textsc{OpenFOAM} v7.0, in combination with the \textsc{RheoTool} v5.0 extension \citep{rheoTool}. Full details of the numerical methodology, including discretisation schemes, solvers, and convergence criteria, are provided in \citet{Sutton-et-al:2022}.

The geometric domain for the flow simulations has the same dimensions and features as the manifold in Fig.~\ref{fig:experimental rig schematic}. To reduce computational cost, we simulated only half of the domain by exploiting a plane of symmetry that passes through the centrelines of all circular channels in the manifold. Flow through the domain is described by the continuity and momentum equations,
\begin{equation}
    \label{eqn:Cont Eqn}
    \hat{\nabla} \cdot \hat{\boldsymbol{u}} = 0 \;,
\end{equation}
\begin{equation}
    \label{eqn:Mom Eqn}
    \rho\left(\frac{\mathrm{\partial}\hat{\boldsymbol{u}}}{\mathrm{\partial}t}+\hat{\boldsymbol{u}}\cdot\hat{\nabla}\hat{\boldsymbol{u}}\right) = -\hat{\nabla} \hat{p} + \hat{\nabla} \cdot \hat{\boldsymbol{\tau}} \;,
\end{equation}
where $\hat{\boldsymbol{u}}$ is the velocity vector, $\hat{p}$ is the pressure, $\hat{\boldsymbol{\tau}}$ is the stress tensor, and $t$ is the time. Though we are ultimately interested in steady-state solutions, a pseudo-transient approach was adopted by maintaining the transient terms in the governing equations. Inertial effects were fully neglected by setting the term $\hat{\boldsymbol{u}}\cdot\hat{\nabla}\hat{\boldsymbol{u}}$ in Eq.~\eqref{eqn:Mom Eqn} equal to zero. Additionally, for stability, we adopted a tensorial formulation of the Herschel-Bulkley constitutive model that includes Papanastasiou regularisation \citep{Papanastasiou:1987, rheoTool}
\begin{equation}
    \label{eqn:computational HB}
    \hat{\boldsymbol{\tau}} = \min{\left(\eta_{\mathrm{max}}, \tau_0 \hat{\dot{\gamma}}^{-1}\left(1-e^{-b\hat{\dot{\gamma}}}\right) + K\hat{\dot{\gamma}}^{n-1}\right)}\hat{\dot{\boldsymbol{\gamma}}} \;\;,
\end{equation}
where $\hat{\Dot{\gamma}}$ is the second invariant of the shear rate tensor, $\hat{\dot{\boldsymbol{\gamma}}}$,
\begin{equation}
    \hat{\dot{\gamma}} = \sqrt{\hat{\dot{\boldsymbol{\gamma}}}:\hat{\dot{\boldsymbol{\gamma}}}/2} \;\;,
\end{equation}
and
\begin{equation}
    \hat{\dot{\boldsymbol{\gamma}}} = \hat{\nabla}\hat{\boldsymbol{u}} + (\hat{\nabla}\hat{\boldsymbol{u}})^{\mathrm{T}} \;\;.
\end{equation}
 In Eq.~\eqref{eqn:computational HB}, $b$ is a regularisation parameter and $\eta_{\mathrm{max}}$ is an upper viscosity bound. We found that using the Papanastasiou regularisation in this study gave increased convergence speeds that were beneficial for such a large domain. The upper viscosity bound eliminates the numerical challenges posed by the viscosity tending to infinity at low shear rates by limiting the viscosity within unyielded regions. We determined that $b=1000$~s and $\eta_{\mathrm{max}}=10^6$~Pa~s were optimal as our results were independent of this parameter for values greater than this. 

A fully developed velocity profile was imposed at the inlet, according to
\begin{equation}
    \hat{u}\left(\hat{r}\right) = \begin{cases}
                \frac{2K}{\hat{P}}\frac{n}{n+1}\left(\frac{\hat{P}}{2K}R - \frac{\tau_0}{K}\right)^{\frac{n+1}{n}} & \mathrm{for}\; \hat{r} < \hat{r}_0 \\
                \frac{2K}{\hat{P}}\frac{n}{n+1}\left[ \left(\frac{\hat{P}}{2K}R - \frac{\tau_0}{K}\right)^{\frac{n+1}{n}} - \left(\frac{\hat{P}}{2K}\hat{r} - \frac{\tau_0}{K}\right)^{\frac{n+1}{n}} \right] & \mathrm{for}\; \hat{r} \geq \hat{r}_0
                \end{cases} \;\;,
\end{equation}
where $R=D/2$ is the circular channel radius, $\hat{r}$ is radial position, $\hat{r}_0$ is the position of the yield boundary, determined by
\begin{equation}
    \hat{r}_0 = \frac{2\tau_0}{\hat{P}} \; ,
\end{equation}
and $\hat{P}$ is the pressure gradient at the inlet pipe. The value of $\hat{P}$ was adjusted to produce the desired bulk flow rate.

The CFD simulations were conducted with no-slip boundary conditions ($\boldsymbol{u}=\boldsymbol{0}$) at the walls, as incorporating wall slip posed significant numerical challenges. As discussed in Section~\ref{section:Predicting Fluid Distribution in a Manifold}, wall slip is negligible under certain flow conditions, and the simulations were used specifically to validate this regime, in line with our objectives: to benchmark the network model and quantify the implications of its assumptions on the flow distribution (fully developed, unidirectional segment flows and neglect of bend/junction dissipation). Wall-slip effects were therefore assessed separately with the experimentally-validated network model.
As additional boundary conditions, we imposed a uniform pressure of zero at the outlets; all other variables had a zero gradient normal to the boundaries.

We blocked individual outlets of the manifold in the experiment and compared the resulting flow distributions to the predictions of the network model. In the CFD simulations, these outlet blockages were replicated by gradually perturbing the outlet boundary conditions until the pressure gradient along the selected branch was zero ($\hat{P} = 0$~Pa~m$^{-1}$). A smooth, time-dependent adjustment of the outlet pressure was found to significantly improve numerical stability compared to an abrupt change.

We used a similar level of mesh refinement to that reported by \citet{Sutton-et-al:2022}, which includes a grid independence study relevant to this class of flows. The used mesh contained 3.89 million elements.

\section{Results and Discussion}
\label{section:results}
In this section, we evaluate the predictions of our network model, utilising the rheological and slip data obtained through independent measurements, against both experimental and CFD data. For clarity, only a representative subset of profiles is shown, focusing on cases with distinct characteristics. We also examine the influence of wall slip on flow uniformity and maldistribution in our manifold system.

\subsection{Predicting Fluid Distribution Profiles in a Manifold}
\label{section:Predicting Fluid Distribution in a Manifold}
\begin{figure}[t]
\centering
\includegraphics[scale=1]{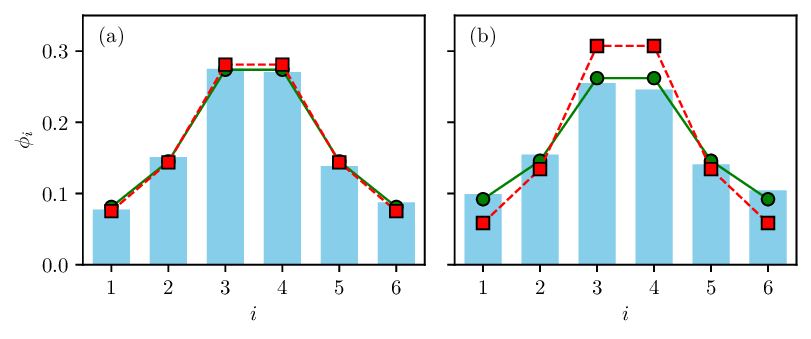}
\caption{Fluid distribution profiles for Carbopol (a) and the emulsion (b) at $B_\mathrm{in}=0.44$. The blue bars indicate the experimental result, red squares indicate the distribution predicted by the network model without slip ($S_1=S_2=0$), and green circles indicate the distribution predicted by the network model when slip is included in the model.}
\label{fig:distribution B=0.44}
\end{figure}

Fig.~\ref{fig:distribution B=0.44} presents the distribution profiles for Carbopol and the emulsion at an inlet Bingham number $B_\mathrm{in} = 0.44$, obtained experimentally and from the network model with and without wall slip. The measured profiles display a slight but consistent asymmetry, with channels~2 and~3 delivering more flow than their nominally symmetric counterparts, channels~5 and~4, respectively. This bias is likely caused by small manufacturing tolerances in the manifold, which alter the hydraulic resistance of specific branches in a repeatable manner.
Despite these minor biases, the agreement between prediction and observation is excellent, with a mean absolute error of 1.1\%. In Fig.~\ref{fig:distribution B=0.44}(a), the predictions with and without slip differ only marginally, indicating that wall slip has little influence on the distribution for this case. In contrast, in Fig.~\ref{fig:distribution B=0.44}(b) accurate prediction of the observed distribution is achieved only when slip is included, resulting in a more uniform profile. This effect is evident from the comparison of the slip and no‑slip model predictions (green and red points, respectively). The influence of wall slip on distribution uniformity in pipe manifolds is examined further in Section~\ref{section:Effect of Wall Slip on the Uniformity of Flow Distribution}.

\begin{figure}[th!]
\centering
\includegraphics[scale=1]{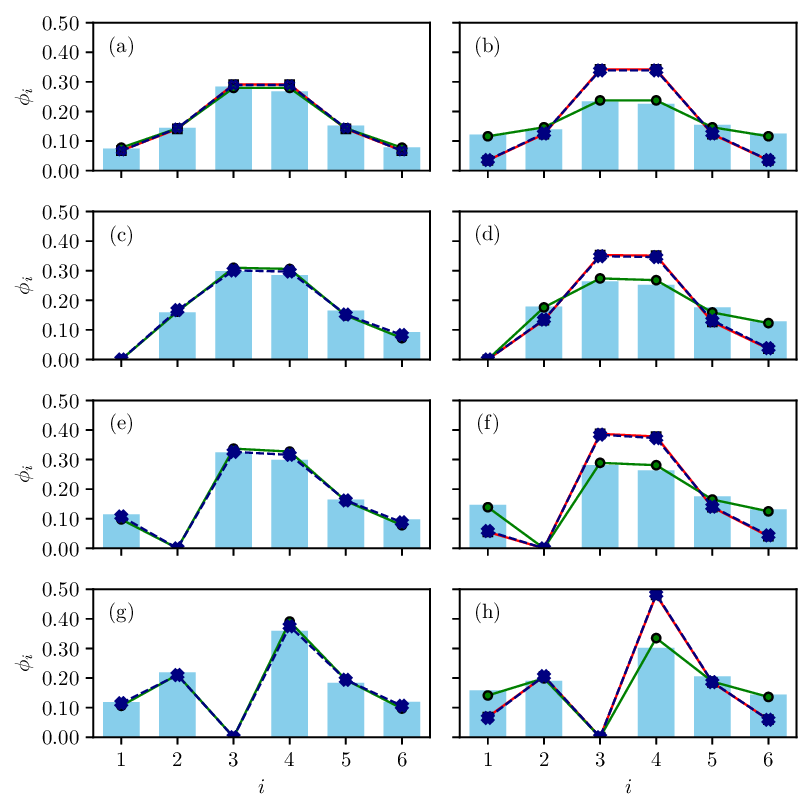}
\caption{Fluid distribution profiles for Carbopol at $B=0.61$ without (a) and with (c, e, g) an outlet blockage and at $B=1.61$ without (b) and with (d, f, h) an outlet blockage. The blue bars indicate the experimental results, the open dark blue crosses indicate the CFD results, the red squares indicate the distribution predicted by the network model without slip ($S=0$), and the green circles indicate the distribution predicted by the network model when slip is included in the model.}
\label{fig:distribution blockages}
\end{figure}

We next change the network's structure by introducing a blockage into an outlet branch of the manifold. We achieve this by plugging the end of a selected outlet for the duration of the experiment. In addition, we attempt to obtain the same distribution from the numerical CFD simulations as the experiment by introducing a perturbation to the pressure boundary condition of the blocked outlet, as described in Section \ref{section:numerical methodology}. We then assess the model's adaptability by attempting to predict the distribution from the new network structure. In the MATLAB code, we introduce a new equation that sets the mass flowrate of the selected outlet to $\dot{m}_{\mathrm{out},i}=0$ and removes the momentum balance equation for this branch in the numerical solution to avoid an overdetermined problem. Fig.~\ref{fig:distribution blockages} shows the distribution profiles at $B=0.61$ and $B=1.61$ with no outlet blocked or with outlets 1, 2, or 3 blocked. We find excellent agreement between the experimental data, CFD data, and the model for all profiles with an inlet condition of $B=0.61$ [Fig.~\ref{fig:distribution blockages}(a, c, e, g)]. The CFD results are particularly close to the network model prediction, with a mean absolute error of 0.18\% for the unblocked case and 0.19\%, 0.10\% and 0.27\% for the cases with blockages in outlet 1, 2, and 3, respectively. In Fig.~\ref{fig:distribution blockages}(b, d, f, h), the experimental distribution profiles do not agree with the CFD profiles because wall slip is present, which does not match the boundary condition imposed in the simulations. However, the experimental profiles match with the prediction of the model when we include slip ($S>0$), and the CFD profiles match with the prediction of the model when slip is absent ($S=0$). The data suggest that the network model can capture the system behaviour seen experimentally and in CFD simulations when introducing outlet blockages, illustrating the adaptability of the model when the network structure is modified. Hence, the accuracy of our model is not simply limited to the manifold design we have used in this study.

Good agreement between the distribution profiles observed experimentally and in the CFD simulations and the network model prediction justifies neglecting minor losses and suggests that more complex rheological behaviour, such as viscoelasticity, of our test fluids do not affect the flow significantly. Without minor losses, the non-uniformity in outlet mass flowrates from our system is solely the result of different resistances to the flow because of different path lengths. Thus, a uniform distribution is obtainable by ensuring the resistance is the same in each branch. It is possible to achieve this by manipulating the pipe length and diameter to give the same pressure drop across every branch, akin to the methodology of \citet{Brod:2003}; however, this would only apply to a given fluid and set of operating conditions, which is useful only if these remain constant. Alternatively, valves introduce additional losses, the size of which is dependent on the opening position. Characterisation of friction losses in valves would allow process operators to predict the required valve position to achieve the desired distribution. Many processing situations are likely to possess viscous-dominated conditions, owing to the large apparent viscosities of many yield-stress materials. However, corrections for minor losses can be included for less viscous fluids, similar to the loss coefficients for Newtonian fluids in \citet{Miller:1978}. 

\subsection{Effect of Wall Slip on Flow Distribution}
\label{section:Effect of Wall Slip on the Uniformity of Flow Distribution}
\begin{figure}[ht!]
	\centering
	\includegraphics[width=1\textwidth]{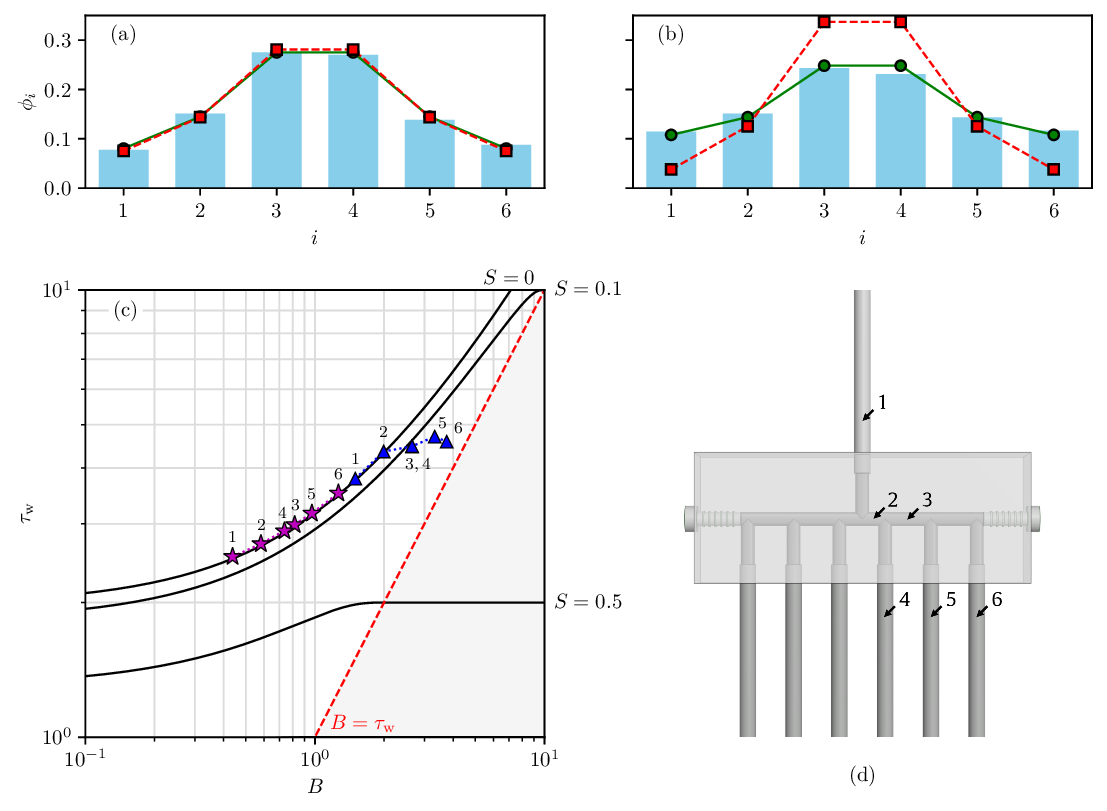}
    \caption{
    Fluid distribution from the six-outlet manifold for Carbopol at (a) $B_\mathrm{in}=0.44$ and (b) $B_\mathrm{in}=1.50$. Blue bars: experiments; red squares: network model without slip ($S=0$); green circles: network model with slip. (c) Predicted wall shear stress, $\tau_\mathrm{w}$, from the network model with slip as a function of $B$ and $S$. The black curves are iso-contours of the slip number $S$; larger $S$ signifies a stronger slip contribution. The red dashed line marks $\tau_\mathrm{w}=B$, which separates the unyielded (plug–slip) regime ($\tau_\mathrm{w}\le B$) from the yielded regime ($\tau_\mathrm{w}>B$). The plot is intended to visualise, for each section, both the relative importance of slip in the transport and the proximity to unyielded conditions. Symbols mark the local $(B,S)$ operating points for the segments labelled in (d), corresponding to the cases in (a) (magenta stars) and (b) (blue triangles) with slip.
    }
    \label{fig:carbopol distributions}
\end{figure}

We now examine the influence of wall slip on the uniformity of fluid distribution from a manifold. Fig.~\ref{fig:carbopol distributions} shows the distribution profiles for Carbopol with $B_{in}=0.44,\;\mathrm{and}\;1.50$. Slip does not affect significantly the distribution in Fig.~\ref{fig:carbopol distributions}(a), but has a clear impact on the distribution in Fig.~\ref{fig:carbopol distributions}(b), shown by comparing the prediction of the network model with a no-slip condition imposed (red line) and the experimental observation. Without correcting for slip, the network model provides an unsatisfactory prediction of the distribution profile for $B_{in}=1.50$ in Fig.~\ref{fig:carbopol distributions}(b). When we include the slip law in the model with the correct constitutive parameters (green line), the prediction is accurate, with a $<2\%$ difference between the prediction and experimental distribution for all outlets. Agreement between the observed distribution profiles and the network model under partial slip conditions demonstrates that the model can accurately capture the slip behaviour in the manifold, provided that we use the correct parameters. In principle, we can invert the problem and use a manifold device as a novel approach to acquire the slip parameters of a material by fitting these parameters to several distribution profiles at different flowrates. 
    Such a device offers a simpler alternative to a capillary rheometer: it requires only the distribution of fluid within the manifold network, not a pressure-drop measurement. Pressure sensing is often unreliable for highly viscous or granular yield-stress materials (e.g. pastes or soft solids), which can block transducer orifices or impede diaphragm actuation.

We can further probe the influence of slip on the flow by plotting the conditions throughout the manifold for these distribution profiles. Fig. \ref{fig:carbopol distributions}(c) shows $\tau_\mathrm{w}$ as a function of $B$ and $S$. The black solid lines show $\tau_\mathrm{w}$ for constant values of $S$, with the topmost line showing $S=0$. The flow is driven solely by slip below the red dashed line ($\tau_\mathrm{w}=B$), and there is no fluid deformation. The pressure drop is independent of $B$ in this regime and is described by Eq.~\eqref{eqn:HB SP Only Slip}. Above the line, both slip and fluid deformation affect pressure drop -- $\tau_\mathrm{w}$ increases as $B$ increases or $S$ decreases. The unyielded plug region occupies a greater proportion of the pipe with larger $B$; therefore, the average apparent viscosity across the cross-section of the pipe and viscous dissipation is greater. Slip has a greater influence on the flow with increasing $S$, reducing friction with the pipe walls. In the limit of $S=0$, the flow is driven solely by fluid deformation with wall conditions described by the no-slip condition. The wall shear stress for this curve approaches $\tau_\mathrm{w} = B$ at high values of $B$. A no-slip wall condition gives the maximum possible friction losses for a given fluid; therefore, there are no friction curves that exist above the $S=0$ curve. The red circles and blue triangles show the local flow condition for Fig.~\ref{fig:carbopol distributions}(a) and \ref{fig:carbopol distributions}(b) in the segments labelled in Fig.~\ref{fig:carbopol distributions}(d). For $B_\mathrm{in} = 0.44$, all segments lie on the $S = 0$ line, indicating that slip is negligible throughout the manifold. At $B_\mathrm{in} = 1.50$, slip remains negligible in the inlet segment, but as the flow divides among branches 3, 4, 5, and 6, the local Bingham number increases, and slip begins to influence the flow. The effect of slip becomes progressively more pronounced with distance from the inlet, with segment 6 exhibiting the highest contribution from wall slip.

Slip produces a more uniform distribution in Fig.~\ref{fig:carbopol distributions}(b) than in Fig.~\ref{fig:carbopol distributions}(a), by reducing the disparity in flow resistance between the outlet branches caused by the differences in path lengths. The contribution of slip to the flow dynamics in outlet branches further from the manifold inlet is always greater than in the inner branches since the flowrate is lower further from the inlet as the flow divides; thus, $S$ is greater in these branches as $S\propto\hat{\bar{u}}^{\beta n-1}$, and the exponent is negative for both of our test fluids. As a result, there is a greater reduction in friction losses in the outlet paths further from the inlet relative to branches closer to the inlet. For weakly shear-thinning fluids, $\beta n-1$ may be positive, and wall slip could have the opposite dependency on the flowrate, likely giving markedly different distribution profiles to those found in this study.

The impact of wall slip on the uniformity of the flow distribution can be quantified over a range of inlet conditions using a maldistribution factor, \(\zeta_{\mathrm{M}}\), defined as
\begin{equation}
    \zeta_{\mathrm{M}} = \sqrt{\frac{1}{N} \sum_{i=1}^N \left(\phi_i - \frac{1}{N}\right)^2},
\end{equation}
where \(\phi_i\) is the fraction of the total flow through outlet \(i\), and \(N\) is the number of outlets. This quantity can be normalised by the symmetry-constrained maximum for this manifold,
\begin{equation}
    \zeta_{\mathrm{M,max}} = \frac{1}{N}\sqrt{\frac{N-2}{2}},
\end{equation}
which corresponds to the case where all the flow is equally divided between the two central outlets (due to symmetry), and the remaining outlets carry no flow. For the six-outlet manifold considered here, \(\zeta_{\mathrm{M,max}} \approx 0.236\). As a result, the normalised maldistribution factor has a value of $0\leq\zeta_{\mathrm{M}}/\zeta_{\mathrm{M,max}}\leq1$ with $\zeta_{\mathrm{M}}/\zeta_{\mathrm{M,max}}=0$ when the output is uniform and the mass flowrate in each branch is identical.

\begin{figure}[t!]
	\centering
	\includegraphics[scale=1]{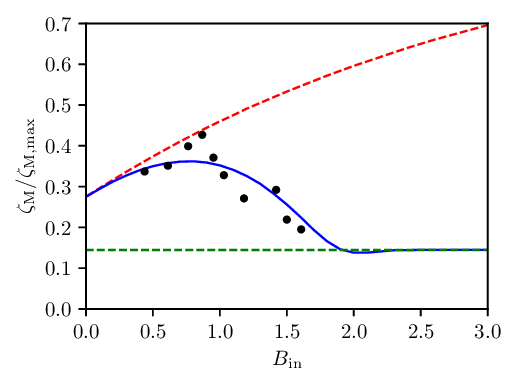}
    \caption{Normalised maldistribution factor ($\zeta_M/\zeta_{\mathrm{M,max}}$) predicted for Carbopol by the network model (blue line) as a function of $B_\mathrm{in}$ and the maldistribution factor of the experimental data (black points). The red-dashed line shows $\zeta_M/\zeta_{\mathrm{M,max}}$ if a no-slip condition applies for all $B_\mathrm{in}$. The green dashed line shows the flow distribution for pure slip in the system.}
    \label{fig:MF carbopol}
\end{figure}

Fig.~\ref{fig:MF carbopol} shows that $\zeta_{\mathrm{M}}$ increases monotonically with $B$ when the no-slip condition (red line, for $S=0$) applies to the entire network. This line represents the asymptotic limit of the distribution at low values of $B_\mathrm{in} \lesssim 0.1$, i.e., for shear stresses significantly larger than the yield stress of the material. As $B_\mathrm{in}$ increases when slip is present, the maldistribution factor decreases, indicating that the flow distribution becomes increasingly uniform due to the effect of wall slip. At low $B_\mathrm{in}$, it is energetically favourable for more material to exit through the innermost branches (3 and 4), which offer the path of least resistance. However, as $B_\mathrm{in}$ increases, wall slip begins to significantly contribute to the flow in the outer branches, reducing their hydraulic resistance relative to the inner branches. This shift in the balance of resistance allows more material to exit through the outer channels. The strengthening of slip effects in the outer branches thus drives the system toward a more uniform distribution profile. The black data points in Fig.~\ref{fig:MF carbopol} show the maldistribution factor \(\zeta_{\mathrm{M}}\) measured experimentally for Carbopol. While the experimental values exhibit noticeable scatter, this can be attributed to small, asymmetries arising from manufacturing tolerances and to uncertainties in the measurement of the output mass from each outlet. Despite this variability, the experimental trend aligns well with the predictions from the network model.

At large inlet Bingham numbers $(B_\mathrm{in} \gtrsim 2)$, the distribution plateaus onto the green curve, which corresponds to the flow distribution under pure wall slip. In this regime, the material is primarily transported in an unyielded state, with the plug region occupying most or all of the channel cross-section. For a given fluid and geometry, this slip-dominated regime is advantageous in that it allows for fluid transport at significantly lower pressure drops compared to yield-driven flow. Moreover, the resulting distribution becomes largely independent of the fluid’s bulk rheological properties and depends instead on tribological and topological interactions at the wall. These properties can, in principle, be tuned by modifying the wall material or surface treatment. However, a potential drawback of this regime is that it may only be accessible at relatively low flow rates, which could limit production throughput. Manipulation of the surface characteristics of the pipe network may expand the range of flow rates that constitute the slip-dominated regime.

These observations illustrate the insights that a maldistribution factor plot such as Fig.~\ref{fig:MF carbopol} can provide, particularly as the uniformity of the distribution profile is often a criterion for the performance of a given manifold design. It allows prediction of the effect of a change in conditions or even system re-design on the manifold's performance. 

\section{Conclusions}
\label{section:conclusions}

We have developed a reduced-order network model that accurately predicts the steady-state distribution of yield-stress fluids in branched pipe manifolds, accounting for wall slip through a power-law slip boundary condition. The model relies solely on major frictional losses and is calibrated using independently measured rheological and slip parameters obtained from rotational and capillary rheometry.

Model predictions were validated against both laboratory-scale experiments and fully resolved CFD simulations, with excellent agreement across a range of flow conditions. The inclusion of wall slip was critical to reproducing the observed distributions, particularly under high Bingham number conditions, where slip significantly reduces flow resistance in distal branches and promotes uniformity.

The model also enables the identification of distinct flow regimes. In particular, we characterise a slip-dominated regime at high Bingham numbers ($B \gtrsim 2$ for the studied geometry), where the fluid is transported primarily through wall slip rather than bulk deformation. This regime allows for lower pressure drops and decouples the flow distribution from the fluid’s bulk rheology, offering potential energy savings and enhanced output uniformity of manifolds in applications where low flow rates are acceptable.

Beyond forward prediction, we propose that the manifold system can be used in reverse: measured distribution profiles can be fitted to infer slip parameters, offering a practical alternative to conventional capillary rheometry that avoids pressure measurements.

This modelling framework is fully predictive, computationally efficient, and readily generalisable to other manifold designs. It requires only fluid property data and geometry as input, making it a valuable tool for the design and optimisation of industrial processes involving viscoplastic materials -- particularly in filling, dosing, and transport operations where uniform distribution and minimal pumping power are critical.

\section*{Acknowledgements}
The authors acknowledge financial support from an EPSRC Prosperity Partnership with Unilever: Centre for Advanced Fluid Engineering and Digital Manufacturing (CAFE4DM) (grant reference EP/R00482X/1). The authors would also like to express appreciation to Martin Quinn for his assistance in fabricating the experimental rig and to Research IT for their assistance and the use of the Computational Shared Facility at The University of Manchester.

\appendix
\section{Model Nondimensionalisation}
\label{appendix:model nondimensionalisation}
We start by considering unidirectional steady flow in a pipe segment is described by the Cauchy momentum equation
\begin{equation}
    \label{eqn:cauchy equation}
    \rho (\boldsymbol{\hat{u}} \cdot \hat{\nabla}\boldsymbol{\hat{u}}) = - \hat{\nabla}\hat{p} + \hat{\nabla}\cdot\boldsymbol{\hat{\tau}}
\end{equation}
We describe the stress response of the materials using the tensorial formulation of the Herschel-Bulkley constitutive model without regularisation
\begin{equation}
\begin{split}
    \label{eqn:HB nondimensionalisation}
    \boldsymbol{\hat{\tau}} = \left(\frac{\tau_0}{\hat{\dot{\gamma}}} + K{\hat{\dot{\gamma}}}^{n-1}\right)\hat{\dot{\boldsymbol{\gamma}}} \hspace{10pt} &\mathrm{for\;} \left|\boldsymbol{\hat{\tau}} \right|> \tau_0 \\
    \hat{\dot{\boldsymbol{\gamma}}} = \boldsymbol{0} \hspace{92pt} &\mathrm{for\;} \left|\boldsymbol{\hat{\tau}} \right| \leq \tau_0 \;\;.
\end{split}
\end{equation}
where $\left|\boldsymbol{\hat{\tau}} \right|$ is the norm of $\boldsymbol{\hat{\tau}}$. We consider the governing equations in nondimensional form, introducing the following nondimensional quantities
\begin{equation}
     r = \frac{\hat{r}}{D} ,\; \boldsymbol{u} = \frac{\boldsymbol{\hat{u}}}{\hat{\bar{u}}} ,\; p = \frac{\hat{p}}{K\left(\frac{\hat{\bar{u}}}{D}\right)^n} ,\; \boldsymbol{\tau} = \frac{\boldsymbol{\hat{\tau}}}{K\left(\frac{\hat{\bar{u}}}{D}\right)^n} \;\;.
\end{equation}
where $r$ is the radial position in the circular channel cross section. Using these nondimensional quantities, Eq.~\eqref{eqn:cauchy equation} becomes
\begin{equation}
    \label{eqn:dimless cauchy equation}
    Re (\boldsymbol{u} \cdot \nabla \boldsymbol{u}) =-\nabla p + \nabla\cdot\boldsymbol{\tau} \;\;,
\end{equation}
and Eq.~\eqref{eqn:HB nondimensionalisation} becomes
\begin{equation}
    \label{eqn:dimless HB 3D}
    \begin{split}
    \boldsymbol{\tau} = \left(\frac{B}{\dot{\gamma}}  + {\dot{\gamma}}^{n-1}\right)\dot{\boldsymbol{\gamma}} \hspace{10pt} &\mathrm{for\;} \left|\boldsymbol{\tau} \right|> B \\
    \dot{\boldsymbol{\gamma}} = \boldsymbol{0} \hspace{81pt} &\mathrm{for\;} \left|\boldsymbol{\tau} \right| \leq B \;\;,
    \end{split}
\end{equation}
where $B$ is the Bingham number as defined in Eq.~\eqref{eqn:Bingham number}.
As the flow between each node of the network is assumed to be a steady fully-developed unidirectional Stokes' flow in a cylindrical pipe driven by a constant pressure gradient $P = \partial p/\partial x$, the term $\boldsymbol{u} \cdot \nabla \boldsymbol{u} = 0$ and Eq.~\eqref{eqn:dimless cauchy equation} simplifies to
\begin{equation}
    \label{eqn:dimless momentum equation}
    \frac{1}{r}\frac{\mathrm{d}\left(r\tau_{rz}\right)}{\mathrm{d} r} = P \;\;,
\end{equation}
and Eq.~\eqref{eqn:dimless HB 3D} to
\begin{equation}
\label{eqn:dimless HB}
\begin{split}
    \tau_{rz} = B + \dot{\gamma}^n \hspace{10pt} &\mathrm{for\;} \tau_{rz} \geq B \\
    \dot{\gamma} = 0 \hspace{40pt} &\mathrm{for\;} \tau_{rz} < B \;\;,
\end{split}
\end{equation}
where $\tau_{rz}$, $\Dot{\gamma}$ and $u_z$ are the only non-zero components of the shear stress tensor $\boldsymbol{\tau}$, rate of strain tensor $\hat{\dot{\boldsymbol{\gamma}}}$, and velocity vector $\boldsymbol{u}$, respectively. Introducing Eq.~\eqref{eqn:dimless HB} into Eq.~\eqref{eqn:dimless momentum equation} yields
\begin{equation}
    B + \Dot{\gamma}^n = \frac{P r}{2}
\end{equation}
for $\tau_{rz} > B$. Rearranging and introducing $\Dot{\gamma} = -\frac{\mathrm{d} u_z}{\mathrm{d} r}$ gives
\begin{equation}
    \label{eqn: vel}
    u_z\left(r\right) = - \int \left(\frac{P r}{2} - B \right)^{\frac{1}{n}}\,\mathrm{d}r \;\;.
\end{equation}
Integrating Eq.~\eqref{eqn: vel} yields
\begin{equation}
    u_z\left(r\right) = - \frac{4}{P}\frac{n}{n+1}\left(\frac{Pr}{2} - B \right)^{\frac{n+1}{n}} + \mathrm{const.} \;\;.
\end{equation}
Applying the boundary condition $u_z\left(r=\frac{1}{2}\right)=u_\mathrm{s}$ gives
\begin{equation}
    u_z\left(r\right) = u_\mathrm{s} + \frac{4}{P}\frac{n}{n+1}\left[\left(\frac{P}{4} - B \right)^{\frac{n+1}{n}} - \left(\frac{P r}{2} - B \right)^{\frac{n+1}{n}}\right] \;\;.
\end{equation}
Introducing the shear stress distribution $\tau_{rz}\left(r\right)=\frac{Pr}{2}$ and wall shear stress $\tau_\mathrm{w}=\frac{P}{4}$, the expression simplifies to
\begin{equation}
    \label{eqn:dimless velocity profile}
    u_z\left(r\right) = u_\mathrm{s} + \frac{1}{\tau_\mathrm{w}}\frac{n}{n+1}\left[\left(\tau_\mathrm{w} - B \right)^{\frac{n+1}{n}} - \left(\tau_{rz}\left(r\right) - B \right)^{\frac{n+1}{n}}\right] \;\;.
\end{equation}
We model the slip velocity as a power-law relationship, dependent on the wall shear stress,
\begin{equation}
    \label{eqn:slip velocity}
    \hat{u}_\mathrm{s} = \alpha{\hat{\tau}_\mathrm{w}}^\beta \;\;.
\end{equation}
Nondimensionalising Eq.~\eqref{eqn:slip velocity} gives
\begin{equation}
    \label{eqn:dimless slip velocity}
    u_\mathrm{s} = S \tau_\mathrm{w}^{\beta} \;\;,
\end{equation}
where $S$ the slip number defined in Eq.~\eqref{eqn:slip number}.
Substituting Eq.~\eqref{eqn:dimless slip velocity} into the velocity profile [Eq.~\eqref{eqn:dimless velocity profile}] results in
\begin{equation}
    u_z \left(r\right) = S \tau_\mathrm{w}^{\beta} + \frac{1}{\tau_\mathrm{w}}\frac{n}{n+1}\left[\left(\tau_\mathrm{w} - B \right)^{\frac{n+1}{n}} - \left(\tau_{rz}\left(r\right) - B \right)^{\frac{n+1}{n}}\right]
\end{equation}
for $\tau_{rz}\geq~B$. Below the yield ($\tau_{rz}<B$), the fluids move at constant velocity
\begin{equation}
     u_z = S \tau_\mathrm{w}^{\beta} + \frac{1}{\tau_\mathrm{w}}\frac{n}{n+1}\left(\tau_\mathrm{w} - B \right)^{\frac{n+1}{n}}\;\;.
\end{equation}
Integrating the velocity profile across the cross-sectional area of the pipe
\begin{equation}
    \frac{4Q}{\pi D^2 \hat{\bar{u}}} = 1 = 8\int^{1/2}_0 u_z r \,\mathrm{d}r \;\;.
\end{equation}
We can split the velocity profile into the plug and fluidised regions
\begin{equation}
    4 r_0 ^2 u_{\mathrm{plug}} + 8\int^{1/2}_{r_0} u_z r \,\mathrm{d}r = 1 \;\;.
\end{equation}
where $r_0 = 2 B/P$ is the radial position that delimits fluidised from non-fluidised regions of the flow. Using integration by parts gives
\begin{equation}
   u_\mathrm{s} - 8\int^{1/2}_{r_0} \frac{r^2}{2}\frac{\mathrm{d} u_z\left(r\right)}{\mathrm{d} r} \,\mathrm{d}r = 1 \;\;.
\end{equation}
Since $\dot{\gamma}=-\frac{\mathrm{d} u_z}{\mathrm{d} r}$,
\begin{equation}
    u_\mathrm{s} + 4 \int^{1/2}_{r_0} \dot{\gamma}r^2\, \mathrm{d}r = 1 \;\;.
\end{equation}
Performing the variable transformation $\mathrm{d}r=\frac{1}{\tau_\mathrm{w}}\mathrm{d}\tau_{rz}$ results in
\begin{equation}
     u_\mathrm{s} + \frac{1}{2\tau_\mathrm{w}^3} \int^{\tau_\mathrm{w}}_{B} \dot{\gamma}\tau_{rz}^2\, \mathrm{d}\tau_{rz} = 1 \;\;.
\end{equation}
Since $\dot{\gamma}=\left(\tau_{rz}-B \right)^{1/n}$,
\begin{equation}
    u_\mathrm{s} + \frac{1}{2\tau_\mathrm{w}^3}\int^{\tau_\mathrm{w}}_{B} \left(\tau_{rz} - B \right)^{1/n}\tau_{rz}^2\, \mathrm{d}\tau_{rz} = 1 \;\;.
\end{equation}
Integrating and rearranging gives, finally,
\begin{equation}
    S \tau_\mathrm{w}^{\beta} + \frac{n\left(\tau_\mathrm{w} - B \right)^{\frac{n+1}{n}}}{2\tau_\mathrm{w}^3}\frac{ \left(n+1\right)\left(2n+1\right)\tau_\mathrm{w}^2 + 2n\left(n+1\right)\tau_\mathrm{w} B + 2n^2 B^2}{\left(n+1\right)\left(2n+1\right)\left(3n+1\right)} = 1
\end{equation}
for $\tau_\mathrm{w}\geq~B$, and
\begin{equation}
    S \tau_\mathrm{w}^{\beta} = 1
\end{equation}
for $\tau_\mathrm{w}<B$.

\section{Wall Slip Characterisation}
\subsection{Capillary Rheometer Validation}
\label{appendix:pressure measurement system}

We validated the accuracy of the pressure measurements in our capillary rheometer by comparing measured pressure drops against analytical predictions for the steady, laminar flow of a Newtonian fluid. Silicone oil (density $\rho = 970$~kg~m$^{-3}$, viscosity $\mu = 1$~Pa~s) was used as the reference fluid. The expected pressure drop for a given flow rate is described by the Hagen–Poiseuille equation
\begin{equation}
    \label{eqn:Hagen-Poiseuille}
    \hat{P} = \frac{128\mu}{\pi D^4} Q
\end{equation}
where $\hat{P} = \Delta \hat{p} / L$ is the uniform pressure gradient along the capillary, and $Q$ is the volumetric flow rate.

\begin{figure}[ht!]
    \centering
    \includegraphics[scale=1]{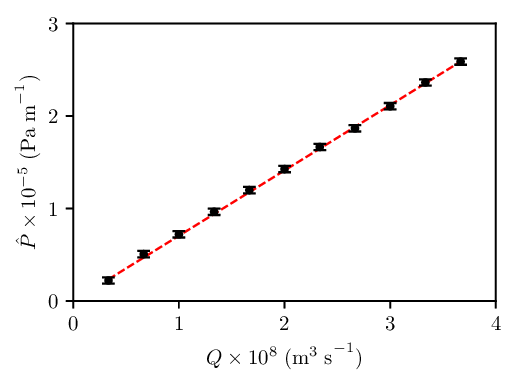}
    \caption{Measured pressure gradient versus flow rate for the flow of silicone oil in the capillary rheometer (black points). The red dashed line corresponds to the theoretical prediction from the Hagen–Poiseuille equation.}
    \label{fig:slip silicone oil}
\end{figure}

Figure~\ref{fig:slip silicone oil} shows that the measured pressure drops are in excellent agreement with theoretical predictions, confirming the validity of the pressure measurement system over the range of flow rates tested.

\subsection{Test Fluid Characterisation}
\label{section:wall slip}
\begin{figure}[t]
\centering
\includegraphics[scale=1]{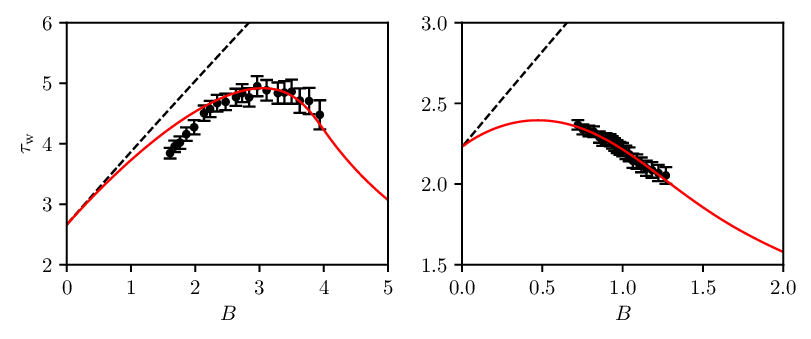}
\caption{Pressure drop measurements of Carbopol (left) and the emulsion (right) across a capillary tube ($D=1.55$~mm) for a range of $B$, presented as the wall shear stress where $\tau_\mathrm{w}=P/4$. The black dashed line indicates the predicted pressure drop in a flow in the absence of slip. The red full line indicates the fitting of Eq.~\eqref{eqn:HB SP Slip} and Eq.~\eqref{eqn:HB SP Only Slip} to the data points with the slip parameters in Table~\ref{tab:material rheology}.}
\label{fig:capillary slip}
\end{figure}

We characterised the wall slip properties of the two test materials using the capillary rheometer setup described in Section~\ref{section:experimental methodology}. Fig.~\ref{fig:capillary slip} shows the wall shear stress for a range of Bingham numbers in the capillary for Carbopol and the emulsion. The wall shear stress was obtained from the measured pressure drop using Eq.~\ref{eqn:pressure drop} and the Bingham number is calculated from the flowrate using Eq.~\ref{eqn:Bingham number}. For the Carbopol gel, the data are better described by a Navier slip law (\(\beta = 1\)), whereas for the emulsion, a quadratic slip law (\(\beta = 2\)) provides a better fit. Below the yield stress, the elastohydrodynamic lubrication theory of \citet{Meeker-Bonnecaze-Cloitre:2004b} predicts a quadratic dependence of slip velocity on wall shear stress (\(\beta = 2\)), arising from deformation of the dispersed phase at the wall. In contrast, \citet{Pemeja-et-al:2019} showed that when viscous dissipation in the interstitial liquid dominates, a linear slip law (\(\beta = 1\)) is observed. For microgels, \citet{Pemeja-et-al:2019} identified a soft transition between these regimes near the yield stress, with \(1 \leq \beta \leq 2\). In emulsions, \citet{Zhang-et-al:2018} reported that the transition from the elastohydrodynamic regime to the interstitial lubrication regime occurs at stresses significantly above the yield stress. The present results are consistent with this picture: the Carbopol gel exhibits linear slip behaviour in the range of imposed stresses, indicative of lubrication dominated by interstitial liquid from stresses close to but lower than the yield-stress, whereas the emulsion shows quadratic behaviour characteristic of elastohydrodynamic lubrication. The slip parameters determined from these measurements are summarised in Table~\ref{tab:material rheology}.

\bibliographystyle{elsarticle-harv} 
\bibliography{references.bib}

\end{document}